\documentstyle[aps,twocolumn,prb,epsfig]{revtex}
\begin{document}

\title{The ground state of the two-leg Hubbard ladder: a
density--matrix renormalization group study} 

\author{ R.M.\ Noack$^a$, S.R.\ White$^b$ and D.J.\ Scalapino$^c$ } 

\address{ a) Institut f\"ur Theoretische Physik, Universit\"at
W\"urzburg, \\ Am Hubland,  97074 W\"urzburg, Germany}

\address{ b) Department of Physics and Astronomy, University of California, 
Irvine, CA, 92717 }

\address{ c) Department of Physics, University of California, Santa
Barbara, CA 93106 }
\date{ \today}

\maketitle
\begin{abstract}
We present density--matrix renormalization group results for the
ground state properties of two--leg Hubbard ladders.
The half--filled Hubbard ladder is an
insulating spin--gapped system, exhibiting a
crossover from a spin--liquid to a band--insulator as a function of
the interchain hopping matrix element.
When the system is doped, there is a parameter range in which the
spin gap remains.
In this phase, the doped holes form singlet pairs and the pair-field and
the ``$4 k_F$'' density correlations associated with pair density
fluctuations decay as power laws, while
the ``$2 k_F$'' charge density wave correlations decay exponentially.
We discuss the behavior of the exponents of the pairing and density
correlations within this spin gapped phase.
Additional one-band Luttinger liquid phases which occur in the large
interband hopping regime are also discussed.

\end{abstract}

\section{\label{INTRO} INTRODUCTION}

There are a number of new materials which contain weakly coupled
arrays of metal--oxide ladders.
For example, SrCu$_2$O$_3$ \cite{hiroi1} and  La$_2$Cu$_2$O$_5$
\cite{cava,nor} contain two--leg ladders with Cu--O--Cu rungs, and 
(VO)$_2$P$_2$O$_7$ \cite{johnston} contains isolated two--leg V--O
ladders.
Alternatively, Sr$_2$Cu$_3$O$_5$ contains three--leg ladders and
Sr$_n$Cu$_{n+1}$O$_{2n+1}$ has $n+1$--leg ladders \cite{hiroi1}.
Magnetic susceptibility and nuclear
relaxation time measurements provide evidence \cite{azuma} that in the
insulating state, the 
even--leg ladders have a spin gap while the odd--leg ladders have
gapless spin excitations.
It is of great interest to understand what happens when holes are
doped into these systems, and, in particular, what happens when holes
are doped into the spin--gapped two--leg ladders.
Recently, Hiroi and Takano \cite{hiroi2} reported that the two--leg
ladder compound La$_2$Cu$_2$O$_5$ could be hole--doped by substituting
Sr for some of the La.
They found that the conductivity of the doped ladders has a metallic
behavior and that magnetic susceptibility data suggests
that a spin gap remains in the lightly doped system.

Here we examine the properties of a Hubbard model on a two--chain
ladder in order to understand the ground state--behavior of the
undoped and doped two--leg ladders in terms of an itinerant electron
model.
We will show primarily numerical results for the energies and equal--time
correlations of the ground and low--lying states obtained using the
Density Matrix Renormalization Group technique \cite{DMRG} (DMRG).
While parts of this work have been published previously
\cite{noack}, here we will discuss the
results in more detail, make comparisons to weak-- and
strong--coupling analytic pictures, and also
present some new results on the local structure of the pairing, the
``$4k_F$'' charge--density--wave (CDW), and on the behavior of the spin and
charge gaps in the quarter--filled system.

The Hubbard and related models on a two--leg ladder have received
much theoretical attention recently, via various analytic
approximations, as well as a variety of numerical techniques.
We direct the reader to a recent review \cite{dagottoreview} and the
references contained therein.
Previous numerical work includes Lanczos calculations for a $t$--$J$
model by Dagotto et al. \cite{dagotto} who suggested that
antiferromagnetic $S=1/2$ coupled chains should have a spin gap and
that if the two--chain system could be doped, it 
would have enhanced superconducting or charge--density--wave
correlations in the ground state.
In Ref.\ 11, 
the Heisenberg model was treated using exact
diagonalization and analytic techniques in order to determine the
ground state properties, the temperature dependence of the spin
susceptibility, and the spin excitation spectrum.
Rice and coworkers
\cite{rice,tsunetsugu} discussed 
a $t$--$J$ model and suggested that if this system were lightly doped, the
spin gap phase would persist and a ground state with dominant
superconducting correlations could be realized.
The one-- and two--particle dynamical correlation functions were
calculated for the half--filled Hubbard ladder using Quantum Monte
Carlo and a Maximum Entropy analytic continuation in Ref.\ 14. 
There have also been a number of weak--coupling
renormalization group calculations 
\cite{varma,schulz,finkel,fabrizio,khveshchenko,balents} which
find evidence for a variety of phases.
We will discuss the results of one of these calculations\cite{balents}
in more detail below and compare them to our numerical calculations.

In the following, we present numerical evidence that indicates
that the half--filled two--leg Hubbard ladder is a spin--gapped
insulator for all non--zero values of the Coulomb interaction $U$, and
the interchain coupling $t_\perp$.
We show that for weak $U$, there is a well--defined spin--liquid
insulator to band insulator transition that corresponds to the $U=0$
one--band to two--band transition.
For stronger $U$, this transition evolves into a smooth crossover.
When the system is doped, the charge gap disappears, but the spin gap
remains for a range of $t_\perp$ associated with overlapping bonding
and antibonding bands.
For $t_\perp$ larger than this range, the numerical results are
consistent with Luttinger liquid behavior.
Within the spin gap phase, we show that there is an attractive pairing
interaction, a $d$--wave--like structure of the pair wave function,
and algebraically decaying pairing and ``$4k_F$'' CDW correlations.
However, the CDW correlations do not seem to decay in the manner
predicted by weak--coupling bosonization treatments of the
system\cite{nagaosa,balents}.

The organization of the paper is as follows:
in section \ref{HUB}, we review the basic properties of the model and
discuss the predictions of weak--coupling renormalization group
and bosonization calculations.
In section \ref{HFHUB}, we exhibit numerical results for the
half--filled system, discuss the properties of the spin liquid state
and the crossover to a band insulator as a function of the
perpendicular hopping.
In section \ref{DOPEDHUB}, we explore the effects of doping holes
into the system, and discuss the phases present as a function of the
perpendicular hopping at various fillings.
We discuss in detail the behavior of the holes within the
gapped spin liquid state that persists as the isotropic system is
doped.

\section{\label{HUB} THE TWO--CHAIN HUBBARD MODEL}

The single band Hubbard model on two coupled chains of 
length $L$ has the Hamiltonian
\begin{eqnarray}
H&=&-t\sum_{i,\lambda\sigma} (c^\dagger_{i,\lambda\sigma}
               c^{\phantom{\dagger}}_{i+1,\lambda\sigma} + h.c.)
    -t_\perp \sum_{i,\sigma} (c^\dagger_{i,1\sigma}
               c^{\phantom{\dagger}}_{i,2\sigma} + h.c.)
\nonumber\\
 && + U\sum_{i\lambda} n_{i,\lambda\uparrow}
 n_{i,\lambda\downarrow}.
\label{hamiltonian}
\end{eqnarray}
Here $c^\dagger_{i,\lambda\sigma}$ and
$c^{\phantom{\dagger}}_{i,\lambda\sigma}$ create and destroy,
respectively, an electron
on rung $i$ and chain $\lambda$ with spin $\sigma$, and
$n_{i,\lambda\sigma} = c^\dagger_{i,\lambda\sigma} c_{i,\lambda\sigma}$.
We will also use the notation 
$n_{i,\lambda} \equiv \sum_\sigma n_{i,\lambda\sigma}$.
The hopping integral parallel to the chains is
$t$, the hopping between the chains $t_\perp$, and $U$ is the on-site
Coulomb interaction.
For the remainder of this work, we set $t=1$ and measure all energies
in units of $t$. 
Within this paper, we will discuss a system with open boundary
conditions both between the two chains and at the ends of the chains.
In other words, site $i,1$ is connected through only one hopping term
to site $i,2$, and site $1,\lambda$ is not directly connected to site
$L,\lambda$.

The non--interacting, $U=0$, Hamiltonian can be diagonalized by
writing the hopping term in terms of bonding and antibonding states on
a rung and fourier--transforming parallel to the chains.
The energy for the infinite two-chain system is then given by
\begin{equation}
\varepsilon_{\bf k} = -(2t \cos k + t_\perp \cos k_\perp)
\label{eqepsilon}
\end{equation}
with ${\bf k}=(k,k_\perp)$,
where $k_\perp=0$ and $k_\perp=\pi$ corresponds to the energy of the
bonding and antibonding band, respectively, and $k$ is the momentum
along the chains.
For $t_\perp=0$, the two bands will be degenerate, and for nonzero
$t_\perp$, they will be separated by an energy $2t_\perp$.

\begin{figure}
\begin{center}
\epsfig{file=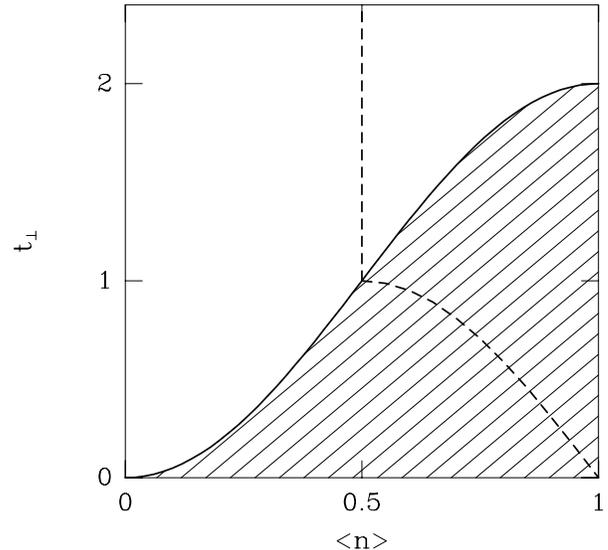, width=8.3cm}
\end{center}
\caption{
The $\langle n \rangle$ -- $t_\perp$ phase diagram for the $U=0$ system.
In the shaded region, both the bonding (lower) and antibonding (upper)
bands are occupied.
At the solid line, the lowest part of the antibonding band just
touches the Fermi surface, and on the dashed line, the bonding
band is half--filled.
}
\label{figu0phase}
\end{figure}

Due to the band structure, the $U=0$ phase diagram in the
$t_\perp$ -- $\langle n \rangle$ plane, shown in 
Fig.\ \ref{figu0phase}, exhibits significant structure.
Both bands will be occupied when $t_\perp < t_{\perp c}$
(shaded region), whereas when $t_\perp > t_{\perp c}$, only the
bonding band will be occupied.
Here $t_{\perp c} = 1 - \cos \pi \langle n \rangle$ and is shown by
a solid line and
$\langle n \rangle \equiv \langle \sum_\sigma c^\dagger_{i,\lambda,\sigma} 
c_{i,\lambda,\sigma}\rangle$ is the band filling.
At half--filling $t_{\perp c}=2$, and the system is 
a two--band metal for $t_\perp < 2$ and a band insulator
with a completely occupied bonding band for $t_\perp > 2$.

\vspace*{-0.8cm}

Along the dashed line in Fig.\ \ref{figu0phase}, the bonding band is
half--filled.
At half--filling ($\langle n \rangle=1$), this occurs at $t_\perp=0$,
where both bands are half--filled.
As $\langle n \rangle$ is reduced, the $t_\perp$ at which the bonding
band is half--filled becomes larger and the occupation of the antibonding
band at this point becomes smaller.
When $\langle n \rangle=0.5$, the bonding band is half--filled only
when the antibonding band is completely unoccupied, i.e. for 
$t_\perp > t_{\perp c}$.

Starting from the band--structure of the noninteracting system, one can
treat the system within a weak--coupling picture by linearizing the
band structure around the Fermi points and taking the continuum limit.
The resulting model can be treated within a renormalization group or
within a bosonization picture.
Balents and Fisher \cite{balents} treated the small $U$, but arbitrary
$t_\perp$ limit in a systematic, controlled procedure. 
In this approach, weak coupling RG
equations were numerically integrated for each value of $\langle n\rangle$
and $t_\perp$, evolving from infinitesimal $U$ to small, but finite
$U$. 
At this point bosonization methods were used to analyze the resulting
Hamiltonians. 
Their weak-coupling results for the phase diagram agree
surprisingly well with our results at rather strong coupling. 
For this reason we will adopt their notation and compare our results
with the results of their approach, which we will denote BFRG. 
An alternative strong--coupling approach, which is less predictive but
perhaps more intuitive, is 
based on a short-range resonating--valence--bond variational ansatz
\cite{rvbanderson,rvbprl}.  
This approach works particularly well in explaining the structure of the
pair wavefunction in the doped, spin-gapped phase. 
We will discuss this in more detail in Section \ref{DOPEDHUB}.

Since, in a weak-coupling picture such as that used by Balents and
Fisher, there are in the general case four Fermi points (at $\pm
k_F^b$ and $\pm k_F^a$), a bosonization treatment yields four possible modes:
two spin modes, one symmetric and one antisymmetric with respect to
interchange of the chains, and two charge modes, one symmetric and one
antisymmetric.
This is a generalization of the bosonization picture for the
one--dimensional Luttinger model, which results in one spin and one
charge mode. 
In general, each of the four modes can either be massive (i.e. gapped),
or massless (with gapless excitations).
Balents and Fisher classify the possible phases according to the
number of massless charge modes $n_c$ and spin modes $n_s$, using a
notation C$n_c$S$n_s$.
In the two--chain system, there are nine possible phases, ranging from
a C0S0 phase with excitation gaps in all four modes to a C2S2 phase,
with all four modes gapless.
For example, the two--chain analog of a Fermi liquid is a C2S2 phase, which
occurs at $U=0$ when both bands are partially filled.
The BFRG calculation, which is valid for weak $U$ but arbitrary
$t_\perp$, finds seven different ground--state phases in the 
$\langle n \rangle$--$t_\perp$ phase diagram.

Since the origin of the BFRG phases can be understood qualitatively in
the context of the $U=0$ phase diagram, we will briefly describe them
here.
At half--filling, the system is in a C0S0 (i.e. spin and charge
gapped) phase for all $U>0$ and $t_\perp>0$.
For $t_\perp < 2.0 $, umklapp processes involving two particles
scattering between the Fermi points lead to gaps opening in all four
modes, producing a spin liquid insulator.
For $t_\perp > 2.0$, the system is a band insulator, as in the
$U=0$ case.
Upon hole doping for $t_\perp > t_{\perp c}$, only the bonding band
comes into play and the behavior is that of a one--band
Luttinger liquid.
Within this phase, there are gapless spin and charge modes, i.e. a
C1S1 phase, except at quarter--filling, where the 
relevant bonding band is half--filled and umklapp processes within
the band lead to a charge--gapped C0S1 phase.
For $t_\perp < t_{\perp c}$ and $\langle n \rangle < 1$, the BFRG
yields a C1S0 phase for most of the phase diagram.
This phase, a spin liquid with one gapless symmetric charge mode, was
also predicted by other, earlier weak--coupling renormalization group
calculations \cite{fabrizio,khveshchenko}, by various
strong--coupling pictures
\cite{rice,gopalan,rvbprl}, and has been found within the Hubbard
and $t$--$J$ models by numerical calculations
\cite{dagotto,tsunetsugu,noack,poilblanc}.
Other phases are found in certain regions where special
processes become relevant.
One such region is in the vicinity of $t_\perp=t_{\perp c}$, 
where the bottom of the antibonding band just touches the Fermi surface
so that its dispersion must be treated quadratically, rather than
linearized.
Here the BFRG calculation finds a C1S0 phase on the band--transition line.
For $t_\perp$ slightly below $t_{\perp c}$, the Fermi velocities in the
bands are quite different and the BFRG finds narrow regions of first
C2S2, then C2S1 phases as $t_\perp$ is reduced.
Umklapp processes become relevant where the bonding band is
half--filled, along the dashed line in Fig.\ \ref{figu0phase}.
Here the BFRG calculation finds a C1S2 phase along this line near
half--filling and near quarter--filling, with an intermediate region
in which the C1S0 phase remains present.

It is interesting at this point to compare the behavior of the
two--chain system to that of the one--dimensional interacting electron gas
(the 1D Hubbard model, for example).
For repulsive short--range interactions, the 1D electron gas is a Luttinger
liquid \cite{haldane}.
A Luttinger liquid has gapless spin and charge excitations (a C1S1
phase), except when umklapp processes
become relevant.
The dominant long--distance behavior of the spin--spin,
density--density, and on--site $s$--wave correlation
functions is power--law decay, and to leading order is \cite{schulz}
\begin{eqnarray}
\langle {\bf S}(r)\cdot {\bf S}(0)\rangle_{\text{TL}} & = & 
r^{-(1+ K_\rho)}\cos(2k_F r) \nonumber \\
\langle n(r) n(0) \rangle_{\text{TL}} & = & r^{-(1+K_\rho)}\cos(2k_F r) 
\nonumber \\
\langle \Delta_s(r) \Delta_s^\dagger(0)\rangle_{\text{TL}} & = &
r^{-(1+1/K_\rho)}
\label{eqtl}
\end{eqnarray}
where ${\bf S}(r)$ is the total spin at site $r$, $n(r)$ is the electron
density, and $\Delta^\dagger_s(r)$ creates an on--site s--wave pair.
The exponent $K_\rho$ is a non--universal parameter dependent on the
system parameters.
There are logarithmic corrections multiplying these expressions which
have been left out for simplicity. 
Umklapp processes become relevant at half--filling for the Hubbard model,
at which point a charge gap develops, leading to a C0S1 mode.
The strong--coupling limit of the half--filled Hubbard model is the
Heisenberg model which has no charge degrees of freedom, and in 1D is
known to have gapless spin excitations.
The spin--spin correlation decays to leading order as \cite{1dheis}
\begin{equation}
\langle {\bf S}(r)\cdot {\bf S}(0)\rangle_{\text{Heis}} = r^{-1} \cos (\pi r).
\label{eqheis}
\end{equation}
In the half--filled Hubbard model, the spin--spin correlation has this
form and other correlation functions decay exponentially.
According to the BFRG calculation, the doped two--chain system should behave
as a Luttinger liquid when $t_\perp > t_{\perp c}$, and the filling of
the Luttinger liquid corresponds to the filling of the bonding band.
Therefore, the correlation functions should decay as power laws
governed by one exponent, as in  Eq. (\ref{eqtl}) in this regime, and
by Eq.\ (\ref{eqheis}) at quarter--filling.

For attractive short--range interactions, the 1D electron gas falls
into the Luther--Emery universality class \cite{lutheremery}.
Within the Luther--Emery model, there is a spin gap and one low--lying
charge mode (C1S0).
The s--wave pairing correlation function and the CDW correlation
function both decay algebraically, with a power parameterized
by the CDW exponent $K_\rho$.
The leading behavior of these two correlation function is given by
\begin{eqnarray}
\langle n(r) n(0) \rangle_{\text{LE}} & = & r^{-K_\rho}\cos(2k_F r) \nonumber \\
\langle \Delta_s(r) \Delta^\dagger_s(0)\rangle_{\text{LE}} & = &
r^{-1/K_\rho}.
\label{eqle}
\end{eqnarray}
Here $K_\rho$ is also dependent on the model parameters, such as the
band--filling, or the strength of the Coulomb interaction, $|U|$, in the
attractive Hubbard model.
Within the BFRG calculation, the doped two--leg ladder falls into a
Luther--Emery--like C1S0 phase for much of the region for 
$t_\perp < t_{\perp c}$.
It is therefore a relevant question whether there are correlations
analogous to those in Eq.\ (\ref{eqle}) with a reciprocal relation
between the exponents in this phase.
Bosonization treatments of the two--chain system
predict\cite{nagaosa,balents} that
the relevant CDW operator will occur at the ``$4k_F$'' wave vector
${\bf q}^* = (2[k_F^b + k_F^a],0)$ with an effective density operator 
$n^{\text{eff}}_i=n_{i,1}n_{i,2}$, and that the pair wave function
will no longer be $s$--wave, but $d$--wave--like with the largest
amplitude across a rung.
Since the behavior of the correlation functions is dependent on the
parameter $K_\rho$, which is a non--universal quantity which can
depend on the parameters of the model, it is important to
determine the exponents within a particular phase.
The least restrictive criterium for the superconducting pairing being
dominant over the CDW is that the correlation function decays more
slowly at long distances, which occurs when $K_\rho > 1$.
However, the consideration of the effects of impurity
scattering and crossover to three dimensions \cite{giamarchi} can lead
one to the more stringent requirement that $K_\rho > 3$, in order to
obtain a stable superconducting state.

The numerical results shown in this work are all calculated using
the DMRG technique \cite{DMRG} on finite lattices of $2 \times 8$ to 
$2 \times 64$ sites.
We obtain the energies and equal--time correlation functions of the
ground state and the low-lying excited states of the finite cluster.
While the DMRG technique gives energies that are, in principle,
variational, it has proven to give quite accurate results for 1D
quantum lattice systems.
The method provides a controlled way of numerically diagonalizing a
finite system within a truncated Hilbert space, by successively
building up part of the system using a real--space blocking
transformation, and then using the reduced density matrix to truncate
the Hilbert space of that part of the system in a controlled way.
One can increase the accuracy by increasing the number of states kept,
and can examine the convergence with the number of states.
Here we typically keep 400 states per block, although 
in the numerically more difficult cases, such as the calculation of
the spin gap on doped $2\times 32$ lattices with small $t_\perp$, we
keep up to 550 states.
Truncation errors, given by the sum of the density matrix
eigenvalues of the discarded states, vary from $3.7\times 10^{-5}$ in
the worse 
case to O($10^{-8}$) in the best cases.
This discarded density matrix weight is directly correlated with the
absolute error in the energy \cite{DMRG}.
It was only possible to obtain accurate calculations on the larger
lattice sizes, $2\times 40$ and $2\times 64$, in certain parameter
regimes, in which the convergence with the number of states was
relatively rapid.
In other regimes, the largest lattice size was $2\times 32$.
We estimate that the maximum errors on the quantities shown in this
paper are at most a few percent and typically are of the order of the
plotting symbol size or less.
We apply open boundary conditions to the lattice because
the DMRG method is most accurate for a given amount of computational
effort with these boundary conditions.

\subsection{\label{HFHUB} Half--filling}

As discussed above, the undoped two--leg ladder materials are gapped
spin--liquid insulators.
Analytic and numerical studies of the Heisenberg model, which is the
strong--coupling limit of the Hubbard model at half--filling, determine
that the system is a spin--liquid insulator for isotropic couplings
\cite{dagotto,barnes,gopalan}.
There is also numerical evidence that this spin--liquid state is
present for {\it all} ratios of the interchain to intrachain coupling
\cite{barnes,rvbprl,otherjp}.
However, it is not clear how the crossover from the weak--coupling
picture dominated by the band structure effects to the
strong--coupling spin--liquid state takes place within the Hubbard model.
We will explore this question in this section.

\begin{figure}
\begin{center}
\epsfig{file=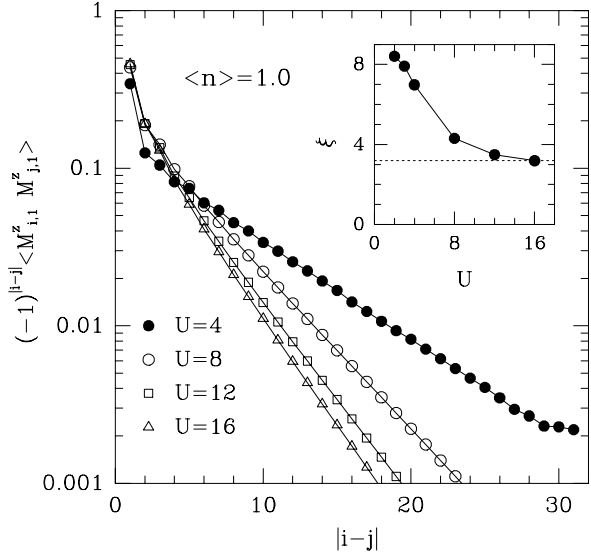, width=8.3cm}
\end{center}
\caption{
The spin--spin correlation function 
$(-1)^\ell \langle M_{i,\lambda}^z M_{j,1}^z \rangle$ for
various $U$ at half--filling ($\langle n \rangle = 1.0$) plotted on a
semilog scale. 
The inset shows the correlation length extracted from the slope of the
lines as a function of $U$, with the dashed line the Heisenberg value
from Ref.\ 23. 
}
\label{figspincorr}
\end{figure}

In order to see the structure of the correlations at half--filling, we
first examine the spin--spin correlation function defined as
\begin{equation}
S(i,j,\lambda)=\langle M_{i,\lambda}^z M_{j,1}^z \rangle
\label{eqSij}
\end{equation}
where $M^z_{i,\lambda} = n_{i,\lambda \uparrow} - n_{i,\lambda \downarrow}$ 
is the $z$--component of the on--site magnetization.
This correlation function, multiplied by a factor $(-1)^i$ which
removes the antiferromagnetic $q=\pi$ structure, is shown in
Fig.\ \ref{figspincorr} on a semilog scale.
Since the correlation functions are straight lines on the semilog
plot, the decay is exponential, as we would expect in a gapped spin
liquid.
The spin--spin correlation function decays more rapidly as $U$ is
increased, implying the correlation length decreases with $U$.
The behavior of the correlation length, obtained by fitting a straight
line to the curves between 6 and 19 lattice spacings, is shown in the
inset.
The dashed line in the inset is the spin correlation length for the
two--chain isotropic Heisenberg system, $\xi=3.19$, taken from 
Ref.\ 23. 
As can be seen, the correlation length for the Hubbard model converges
to the Heisenberg value for large $U$, as one would expect.

We have examined the
spin--spin correlation function for a variety of $U$ and $t_\perp$ at
half--filling, down to $t_\perp=0.2$ at $U=1$, and $U=0.5$ at
$t_\perp=1.0$. 
The decay of the correlation functions is consistent with
exponential decay and a finite correlation length at half--filling for
all $t_\perp$ and $U$ we have examined.
Therefore, we believe the spin liquid phase is present at all $t_\perp$ and
$U$ at half--filling.
This is consistent with weak--coupling renormalization group
treatments \cite{khveshchenko,balents}.

\vspace*{-0.8cm}

We have also examined other correlation functions, such as the
density--density and pair correlation functions, for the
half--filled system.
In all cases they also decay exponentially, with a correlation length
shorter than that of the spin--spin correlation function.

\begin{figure}
\begin{center}
\epsfig{file=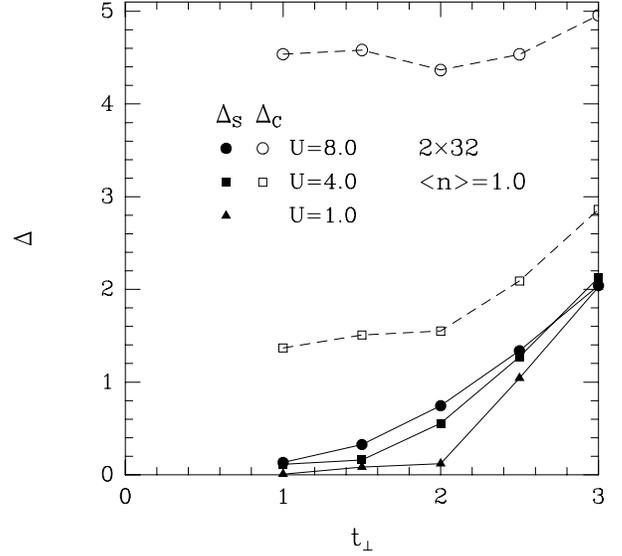, width=8.3cm}
\end{center}
\caption{
The spin gap $\Delta_s$ and the charge gap $\Delta_c$ as a function of
$t_\perp$ at half-filling
on a $2 \times 32$ lattice for various $U$.
}
\label{figgapshalf}
\end{figure}

We next consider the charge and spin gaps defined by
\begin{eqnarray}
\Delta_C  & = &\left[ E_0(N-1,N-1) + E_0(N+1,N+1) \right . \nonumber \\
          && - \left .  2\,E_0(N,N)\right] \big/ 2
\end{eqnarray}
and
\begin{equation}
\Delta_S = E_0(N+1,N-1) - E_0(N,N),
\end{equation}
respectively.
Here $E_0(N_\uparrow,N_\downarrow)$ is the ground state energy for
$N_\uparrow$ spin-up electrons and $N_\downarrow$ spin-down electrons.
Since our current DMRG program cannot target states of a particular
transverse symmetry, we can only calculate the gaps to the
lowest--lying spin and charge excited states, irrespective of symmetry.
We show spin and charge gaps for a range of $U$ and
$t_\perp$ in Fig.\ \ref{figgapshalf}.
For $U=0$, the system is a two--band metal, as discussed in section
\ref{HUB}, for $t_\perp < 2$, and a band insulator for $t_\perp > 2$. 
Therefore, for $U=0$ both gaps would be zero for $t_\perp < 2$, and
would be set by the band separation $2(t_\perp-1)$ for $t_\perp > 2$.
For the $U=1$ spin gap in Fig.\ \ref{figgapshalf}, the behavior for
$t_\perp > 2$ follows the $U=0$ form.
However, for $t_\perp < 2$, we expect a nonvanishing spin gap due to
the spin liquid state, consistent with the exponentially decaying
spin--spin correlation function.
We find numerically that the gap in this region is finite, but relatively
small, with its size associated with the strength of the antiferromagnetic
correlations in the spin liquid state.
We calculate the spin gap over a limited range of $t_\perp$ because
the numerical accuracy of the DMRG method and the size of the spin gap
both become smaller with lower $t_\perp$.
Since one needs to calculate small energy differences and do finite
size scaling in order to accurately determine the spin gap, we find
that an examination of the ground state spin--spin correlations, discussed
above, is a more sensitive test for the existence of the spin liquid
state, given a particular numerical accuracy.
We omit the charge gap for $U=1$ because the accuracy of the
calculated charge gap is comparable to its size in this region.

\vspace*{-0.8cm}

As $U$ is increased, the relatively sharp transition from a spin
liquid insulator to a band insulator becomes a smooth crossover.
This can be understood in terms of the spin gap because in the
Heisenberg limit, the size of the spin gap is set by the
perpendicular coupling $J_\perp \approx 4 t_\perp^2/U$ for large
$J_\perp$.
One can see the crossover to this behavior for $U=4$ and 8, and
intermediate $t_\perp$.
However, as $t_\perp$ becomes larger, the Heisenberg mapping breaks
down (when $4 t_\perp^2$ is of the order of $U$), and the linear
growth in $t_\perp$ characteristic of the band insulator is restored.
For large $U$, $\Delta_c \approx U/2$ at half--filling, and we can see
that the charge gap is approximately this size for $U=4$ and 8.
For $U=4$, the crossover to linear growth of the charge gap with
$t_\perp$ is visible for $t_\perp > 2$.

\begin{figure}
\begin{center}
\epsfig{file=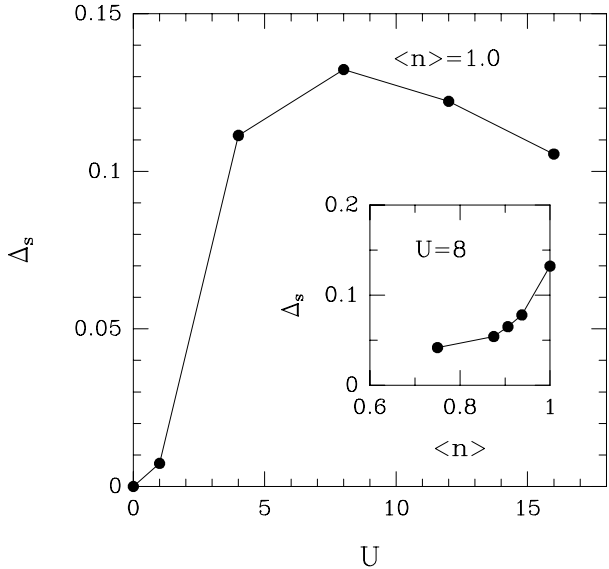, width=8.3cm}
\end{center}
\caption{
The spin gap $\Delta_s$ as a function of $U$ at half-filling on a 
$2 \times 32$ lattice.
The inset shows $\Delta_s$ as a function of filling for $U=8$.
}
\label{figgapu}
\end{figure}

We examine the dependence of the spin gap on $U$ at half--filling
for the isotropic ladder ($t_\perp=1.0$) in Fig.\ \ref{figgapu}. 
For small $U$, the spin gap increases with $U$ as one might expect
from a weak--coupling picture.
For very large $U$, the spin gap should scale with 
$J \approx 4 t^2/U$, since $J$ is the only energy scale in the
Heisenberg system.
Therefore, the spin gap will have a peak at some intermediate $U$, seen
to be approximately $U=8$ in Fig.\ \ref{figgapu}.

\vspace*{-0.8cm}

Therefore, at half--filling, the two--leg Hubbard ladder is
an insulator with a spin and charge gap for all $U$ and $t_\perp$,
with a crossover from a spin--liquid insulator to a band insulator at
$t_{\perp c} =2.0$ that grows less sharp as $U$ increases.

\subsection{\label{DOPEDHUB} The Doped System}

The evolution of the spin liquid state of the half--filled system with
doping is a quite interesting question.
As discussed above, weak--coupling treatments such as the BFRG
described above make rather specific predictions about the phase
diagram of the doped system.
In particular, it is expected that when the isotropic system is doped,
the spin gap will remain, there will be one low--lying charge mode,
and that there will be pairing of the charge carriers.
In addition, according to the BFRG picture, there will a set of phase
transitions associated with the weak--coupling one--band to two--band
transition, and with the umklapp processes that occur when one of the
bands is half--filled.
It is not clear to what extent this picture will hold within a Hubbard
model on a lattice and with a realistic band structure, or at the
relatively strong coupling, $U=8$, treated here.
The detailed properties of the system within each phase, such as the
non--universal 
exponents that determine the relative strength of various correlation
functions, have not yet been calculated in detail using weak--coupling
methods and are thus interesting to determine.

One can also approach the behavior of this system from a strong
coupling point of view.
This has been done within the $t$--$J$ model by starting from the limit of
strong magnetic coupling between the chains\cite{rice}, $J_\perp >> J$.
One can then treat the interactions between the rungs perturbatively,
and, for small doping, make states that are delocalized products of
simple excitations of the rungs.
This gives a physical picture of the behavior for a low density of
holes, and predicts pairing.
In Ref.\ 13, 
it is then argued that the behavior at isotropic
coupling is continuously connected to that at large large $J_\perp$ by
showing that, within exact diagonalization calculations, there is a
continuous evolution of the properties with $J_\perp$, 
and a qualitative correspondence
between the large $J_\perp$ case and the isotropic case.
A variational dimer resonating--valence--bond (RVB) state
\cite{rvbprl} used to treat
the Heisenberg ladder gives a qualitatively similar picture,
and helps justify the continuity between small and large $J_\perp$.
In the following, we will examine the properties of the model as a
function of
filling, and then take slices through the phase diagram at constant 
$\langle n \rangle$ and varying $t_\perp$ at two fillings: 
$\langle n \rangle = 0.875$ and $\langle n \rangle = 0.5$ 
(quarter filling). 

We first examine the behavior of the spin gap and the spin--spin
correlation function when the half--filled system with isotropic
coupling ($t_\perp=1.0$) is doped with holes.
Upon doping, the spin gap, as shown in the inset of 
Fig.\ \ref{figgapu} for a $2\times 32$ lattice, is reduced but remains 
finite down to a filling of at least $\langle n \rangle =0.75$.
Although these results are calculated on a finite lattice, we will
examine the behavior as a function of system size and extrapolate to
the thermodynamic limit at particular fillings below.
In the regime shown in the inset of Fig.\ \ref{figgapu}, the spin gap
remains finite in the thermodynamic limit.

\begin{figure}
\begin{center}
\epsfig{file=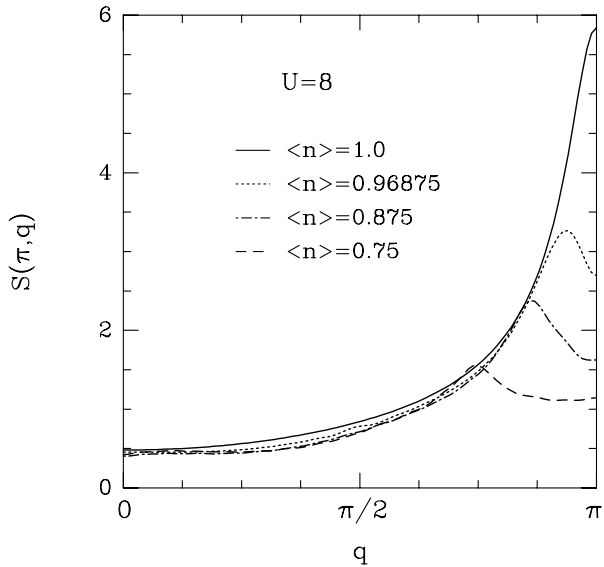, width=8.3cm}
\end{center}
\caption{
The continuous Fourier transform $S(q,\pi)$ of
the spin--spin correlation function $S(i,j,\lambda)$ for $U=8$ at various
fillings on a $2 \times 32$ lattice.
}
\label{figSq}
\end{figure}

The Fourier transform of the ${\bf q} = (q,\pi)$ branch of the
spin--spin correlation function is shown 
in Fig.\ \ref{figSq} for fillings between $\langle n \rangle = 0.75$
and $\langle n \rangle = 1.0$.
In order to reduce the effects of the broken translational invariance
due to the open boundary conditions,
we average the spin--spin correlation function over up to six $i,j$
pairs for each $|i-j|$ in real space before fourier--transforming.
A continuous range of frequencies is shown because the correlation function
decayed sufficiently rapidly so that we could obtain its value for all 
non-negligible separations. 
Thus the actual lattice size used in the
calculation is irrelevant, and unlike the usual treatment of a finite system
with periodic boundary conditions, a straightforward Fourier transform
yields valid results at any frequency value, provided that the correlation 
function is assumed to vanish at larger separations. 
The same approach could also be applied to calculations imposing
periodic boundary conditions, provided that the correlation function
decays to a value sufficiently close to zero for separations greater
than half the system size.
We show only the $q_\perp=\pi$ branch because the correlations are
predominantly antiferromagnetic between the chains and the $q_\perp=0$
branch is thus small and flat, with no interesting features.
At half--filling ($\langle n \rangle = 1.0$), the Fourier transform
$S(q,\pi)$ peaks at 
$q=\pi$ and has Lorentzian
line shape, as one would expect for
a spatially alternating, exponentially decaying function.
As the system is doped, the peak shifts to smaller values of $q$
commensurate with the lattice filling, so that it occurs at the
``$2k_F$'' wave vector $q^* = k_F^b + k_F^a = \langle n \rangle \pi$.

\vspace*{-0.8cm}

\begin{figure}
\begin{center}
\epsfig{file=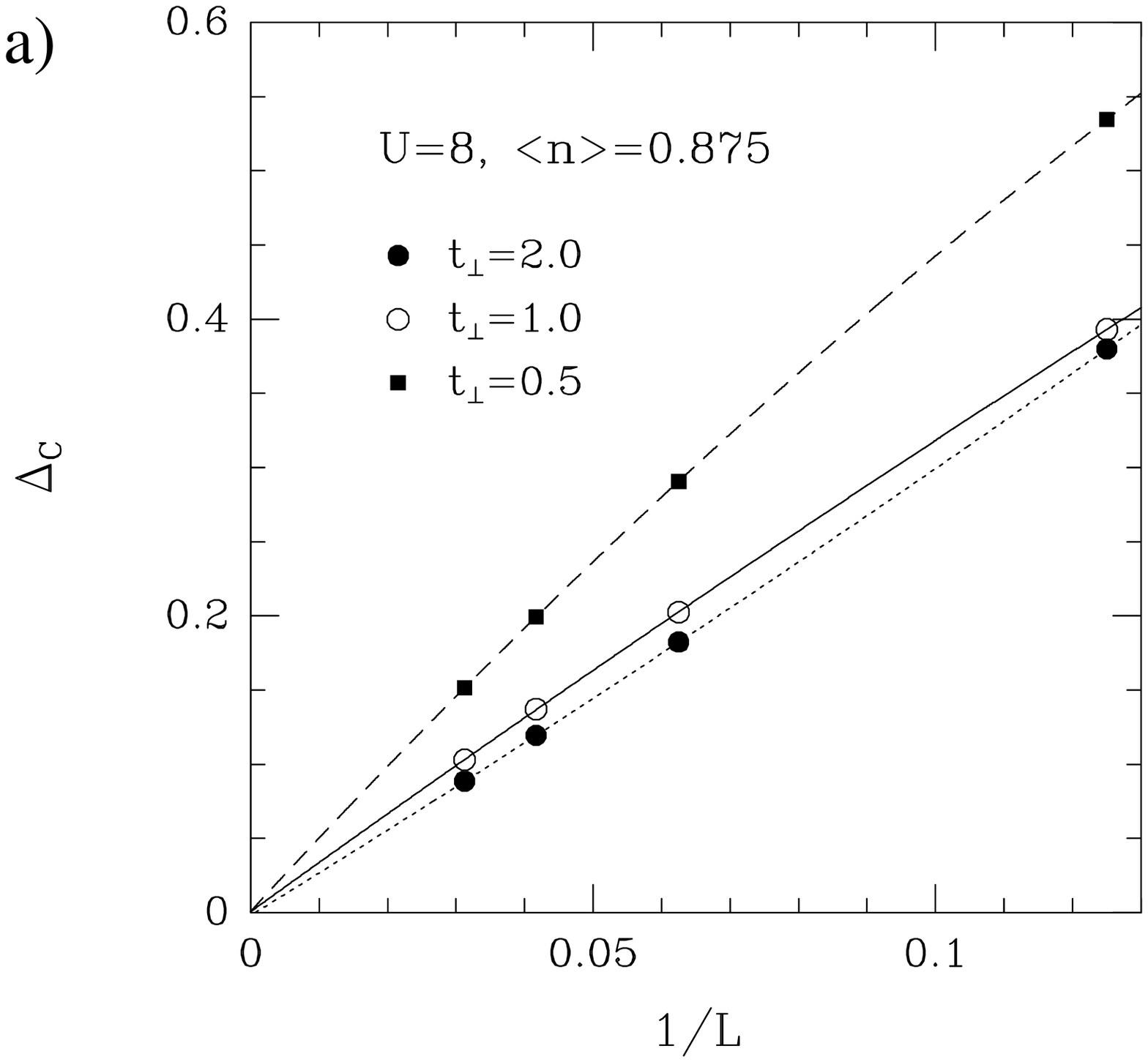, width=8.3cm}
\epsfig{file=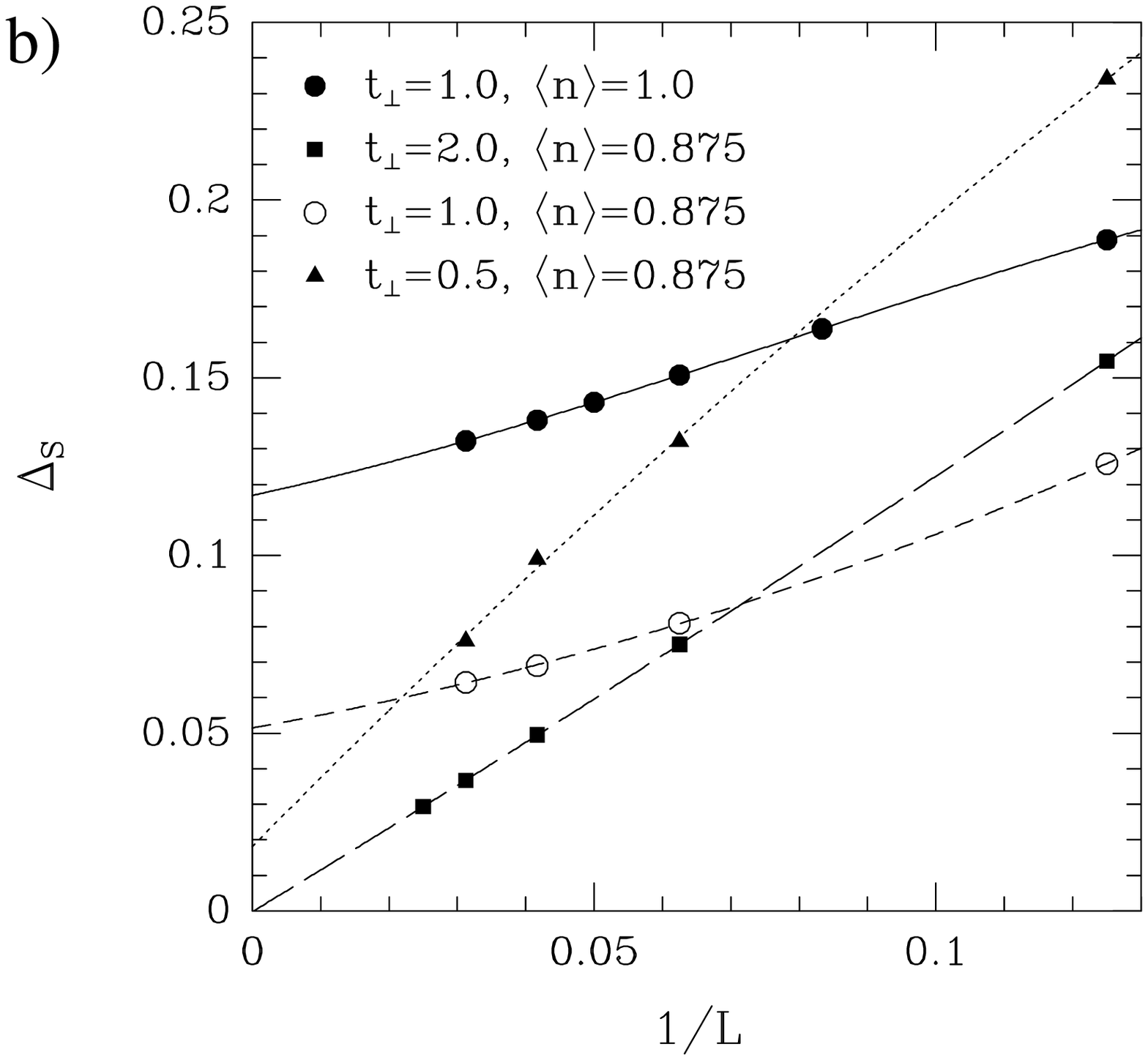, width=8.3cm}
\end{center}
\caption{
(a) The charge gap
$\Delta_c$ and (b) the spin gap $\Delta_s$ for $U=8$ and 
$\langle n \rangle = 0.875$, as a function of the
inverse chain length, $1/L$.   The spin gap at half--filling is shown
for comparison in (b), and the lines are least--squares fits of
polynomials to the data, as explained in the text.
}
\label{figgapsl}
\end{figure}

We have found that the finite--size effects for the charge and spin
gaps can be large for a system with open boundary conditions, even on
the largest lattice, $2\times 32$ sites.
We therefore must examine the gaps as a function of system size and
extrapolate to the thermodynamic limit.
We concentrate first on a particular filling, $\langle n \rangle=0.875$,
relatively close to half--filling.
The charge gap $\Delta_c$ and the spin gap, $\Delta_s$ are plotted as
a function of $1/L$ for three representative $t_\perp$ on lattice
sizes up to $2\times 40$ in Fig.\ \ref{figgapsl}.
The charge gap, Fig.\ \ref{figgapsl}(a), scales approximately
linearly with $1/L$ for all three values of $t_\perp$.
The lines are least--squares fits to polynomials in $1/L$ whose order
is chosen so that the number of free parameters is one less than the
number of $L$ points.
The $1/L\rightarrow 0$ extrapolated value is zero to within the accuracy of
the extrapolation in all three cases.
We believe that this behavior is representative and that the charge gap
vanishes for all $t_\perp$ and $U$ at this filling.
The spin gap, Fig.\ \ref{figgapsl}(b), shows different scaling
behavior for $t_\perp=0.5$, $1.0$, and $2.0$.
For $t_\perp=2.0$, the scaling is approximately linear in $1/L$ and
the gap goes to zero in the thermodynamic limit.
For $t_\perp=1.0$, the $1/L^2$ term is substantial, and 
$\Delta_s(1/L\rightarrow 0) \approx 0.05$.
For $t_\perp=0.5$, the finite--size corrections are larger, the linear
term in $1/L$ is large, and the $1/L^2$ coefficient is negative.
Here $\Delta_s(0) \approx 0.02$, a small but finite value.
However, the uncertainty is large because
finite--size corrections are large and because the errors in $\Delta_s$ at
$t_\perp=0.5$ are relatively large.
(In coupled chain systems, the accuracy of DMRG is generally inversely
related to the coupling between chains.)
Also shown for comparison is the spin gap for the half--filled system at
$t_\perp=1.0$, for which $\Delta_s(0) \approx 0.12$.

\vspace*{-0.8cm}

\begin{figure}
\epsfig{file=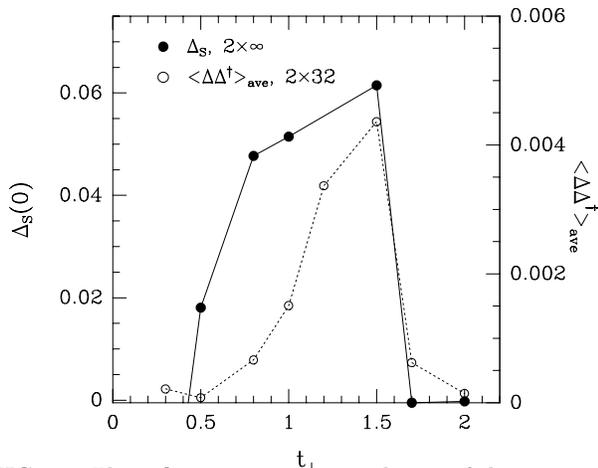, width=7.5cm}
\caption{
The infinite system extrapolation of the spin gap, $\Delta_s(0)$, 
from Fig.\ 6(b) for 
$\langle n \rangle = 0.875$ and $U=8$, and the average of the
pair--pair correlation function 
$\langle \Delta(i) \Delta^\dagger(j) \rangle$
averaged between $|i-j| = 8$ and 12 lattice spacings.
}
\label{figgapinf}
\end{figure}

The extrapolated $1/L\rightarrow0$ values of the spin gap,
$\Delta_s(0)$, are shown 
plotted versus $t_\perp$ in Fig.\ \ref{figgapinf}.
For $t_\perp > 1.7$, $\Delta_s(0) = 0$.
For $U=0$, the one band to two--band transition takes place
at $t_\perp = 1.85$ at $\langle n \rangle = 0.875$.
Within the BFRG picture \cite{balents}, the system should
be a one--band Luttinger liquid, i.e. have no spin gap and charge gap
for $t_\perp>t_{\perp c}$.
The BFRG also predicts some additional phases when $t_\perp$ is
slightly below
the band transition point where the Fermi velocities in the two bands
are very different, and at $t_\perp = t_{\perp c}$, when the
antibonding band can no longer be treated by linearizing the Fermi
surface.
The additional phases below $t_{\perp c}$
have gapless spin and charge excitations, so the
disappearance of the spin gap at a $t_\perp$ somewhat less than
$t_{\perp c}$ is consistent with the existence of these phases,
although we cannot distinguish between symmetric and antisymmetric
spin and charge excited states within our DMRG calculation.
We have, however, not yet seen any evidence for a reappearance of the
spin gap exactly at $t_\perp=t_{\perp c}$.

\vspace*{-0.8cm}

\begin{figure}
\begin{center}
\epsfig{file=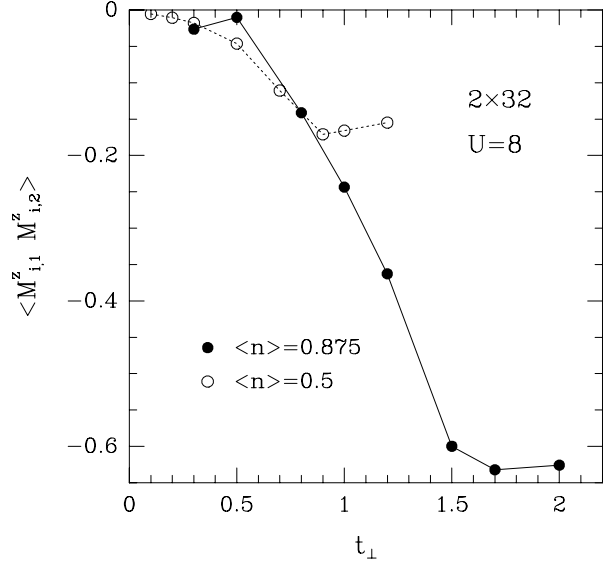, width=8.3cm}
\end{center}
\caption{
The nearest--neighbor spin--spin correlation measured across a rung, 
$\langle M_{i,1}^z M_{i,2}^z \rangle$, at $\langle n \rangle = 0.875$
and $\langle n \rangle = 0.5$ as a function of $t_\perp$ for $U=8$.
Here we have chosen $i=16$ on a $2 \times 32$ lattice.
}
\label{figspincross}
\end{figure}

The infinite system extrapolation of the spin gap, $\Delta_s(0)$,
becomes zero at
$t_\perp \approx 0.38$ and negative for smaller $t_\perp$,
although $\Delta_s$ is positive for all the finite lattices we have
examined.
In addition, the nearest--neighbor spin--spin correlation 
$\langle M_{i,1}^z M_{i,2}^z \rangle$ goes to zero and has a
discontinuity in slope at $t_\perp \approx 0.5$, as shown in Fig.\
\ref{figspincross}.
For $U=0$, the bonding band is half--filled at $t_\perp= 0.4$, and
the BFRG picture predicts that Umklapp processes in the bonding band
will become relevant at this point, leading to a C1S2 phase,
consistent with a vanishing spin and charge gap.
Therefore, the vanishing of $\Delta_s(0)$ at $t_\perp \approx 0.38$ could
be associated with this weak--coupling feature.

For $t_\perp < 0.5$, the finite size effects become large and the
numerical accuracy of the DMRG procedure is reduced, so we have not
been able to unambiguously determine whether or not the spin gap
reappears in the small $t_\perp$ phase at this filling, as would be
predicted by the BFRG.

\vspace*{-0.8cm}

\begin{figure}
\begin{center}
\epsfig{file=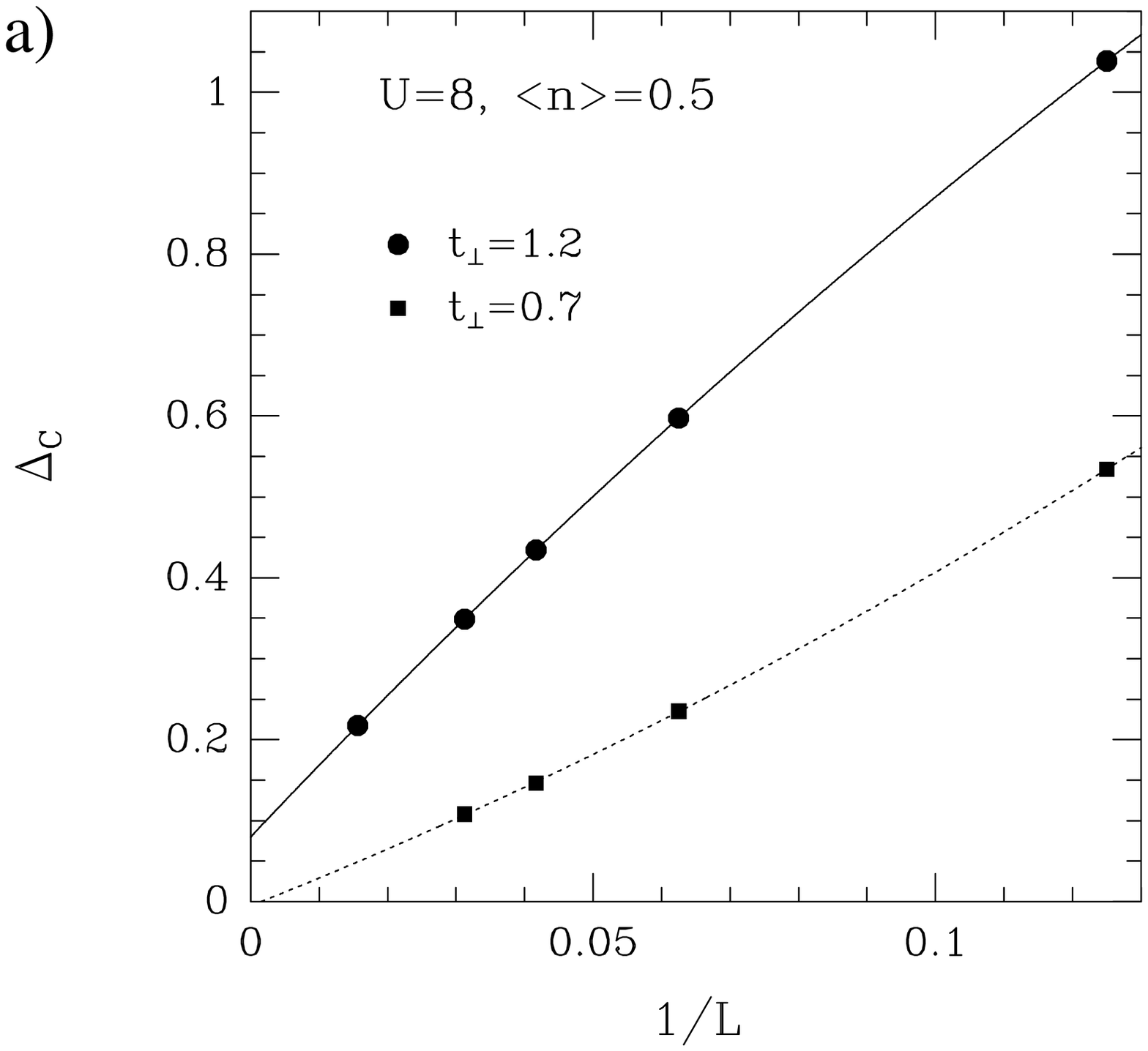, width=8.3cm}
\epsfig{file=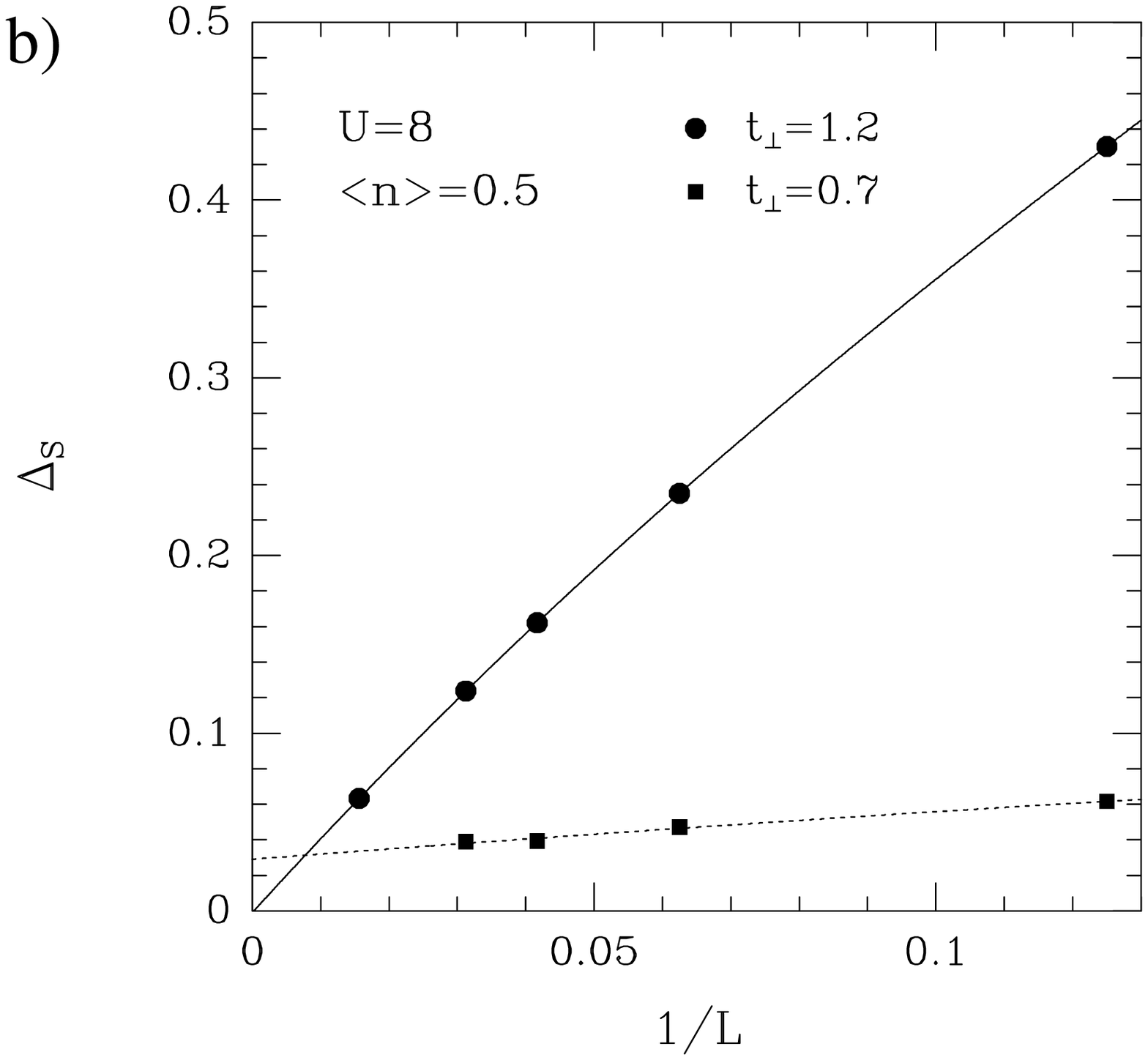, width=8.3cm}
\end{center}
\caption{
(a) The charge gap $\Delta_c$ and (b)
the spin gap, $\Delta_s$, plotted as a function of $1/L$ for $U=8$,
$\langle n \rangle =0.5$, and for 
$t_\perp=1.2$ and 0.7.
}
\label{figgapsln5}
\end{figure}

For the quarter--filled system, the $U=0$ one--band to two--band
transition occurs at $t_\perp=1.0$.
For $t_\perp > 1.0$, the bonding band is half--filled, while for
$t_\perp < 1.0$, both bands are fractionally filled.
The BFRG predicts that within the one--band regime, the system will
behave as a half--filled Luttinger liquid, i.e. a C0S1 phase, for
which $\Delta_c \ne 0$ and $\Delta_s = 0$.
We exhibit the charge and spin gaps as a function of $1/L$ in 
Fig.\ \ref{figgapsln5}.
For $t_\perp = 1.2$, $\Delta_c(0) \approx 0.08$, whereas
$\Delta_s(0)$ vanishes to within the accuracy of the
extrapolation.
Since the finite size effects are fairly large, we have studied
lattice size up to $2\times 64$.
This is possible for $t_\perp=1.2$ since the
numerical accuracy is high at this parameter range, with a maximum
discarded density matrix weight of $6\times 10^{-7}$, keeping 350
states on the $2\times 64$ lattice.
For $t_\perp=0.7$, which is in the two--band region for $U=0$,
$\Delta_c(0)$ vanishes and $\Delta_s(0) \approx 0.07$.
This behavior is consistent with the C1S0 phase predicted by BFRG in this
region.
In addition, since the bonding band is always less than half--filled
for $t_\perp <t_{\perp c}$ at quarter filling, the BFRG calculation
predicts no additional phases due to umklapp effects for small
$t_\perp$.
As can be seen in Fig.\ \ref{figspincross}, the behavior of the
spin--spin correlation across a rung, 
$\langle M_{i,1}^z M_{i,2}^z \rangle$, is smooth in the small
$t_\perp$ region, unlike at $\langle n \rangle = 0.875$, consistent
with the BFRG prediction.

We now concentrate on the doped spin gap phase (C1S0), present for isotropic
$t_\perp$, and small to moderate doping.
This phase could be relevant for  the La$_{1-x}$Sr$_x$CuO$_{2.5}$
materials if, in fact, the ladders are sufficiently weakly 
coupled \cite{LaSrnote}.
The general behavior of holes doped into a spin liquid phase 
might also be relevant to the high--$T_c$ materials.
In particular, we are interested in whether there is an effective
attractive potential between the holes, the nature of the state that
they form, and the type of correlations dominant at long distances.

\vspace*{-0.8cm}

\begin{figure}
\begin{center}
\epsfig{file=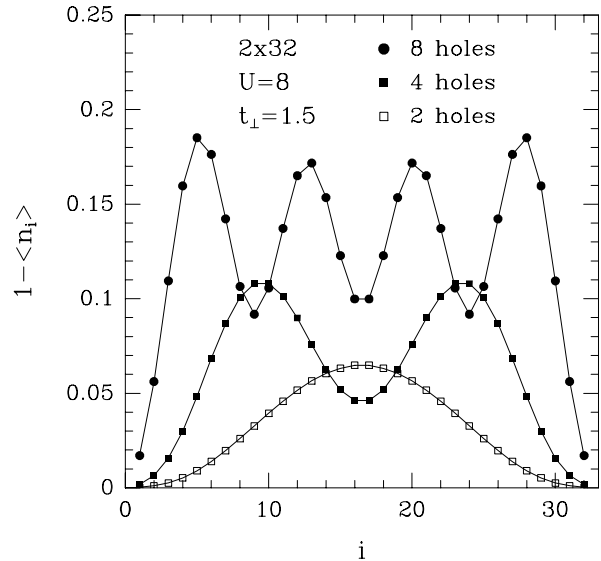, width=8.3cm}
\end{center}
\caption{
The local hole density $1-\langle n_i \rangle$ as a function of
position along the chain $i$ for 2, 4, and 8 holes 
($\langle n \rangle = 0.96875, 0.9375, 0.875$) on a $2\times 32$ chain
for $U=8$ and $t_\perp=1.5$.
}
\label{figholeden}
\end{figure}

We first examine the local behavior as pairs of holes are doped into
the system.
In Fig.\ \ref{figholeden} we exhibit the local hole density 
$1-\langle n_i \rangle$ plotted as a function of the rung index $i$.
Since the system is symmetric with respect to exchange of the chains,
we calculate the average value on a rung.
The inhomogeneous structure is due to the open boundary conditions as
well as the interaction between the holes.
Due to kinetic energy effects, the boundaries repel the holes.
When two holes are put into the system, as can be seen in 
Fig.\ \ref{figholeden}, the hole density has one peak with a maximum
at the center of the ladder.
In other words, the holes tend to both be near the same rung of the ladder,
indicating a net attraction between the holes.
We have calculated the pair binding energy of two holes in a half--filled
$2\times L$ lattice,
defined as
\begin{equation}
E_b = 2E_0(L,L-1) - E_0(L,L) - E_0(L-1,L-1)
\end{equation}
on a $2\times 32$ lattice with $t_\perp=1.0$, and obtain
a positive binding energy, $E_b\approx 0.14$.
As more holes are added to the system, the number of peaks in the
hole density is equal to the number of hole pairs, indicating
that there is a tendency for the hole pairs to repel one another.

\vspace*{-0.8cm}

\begin{figure}
\begin{center}
\epsfig{file=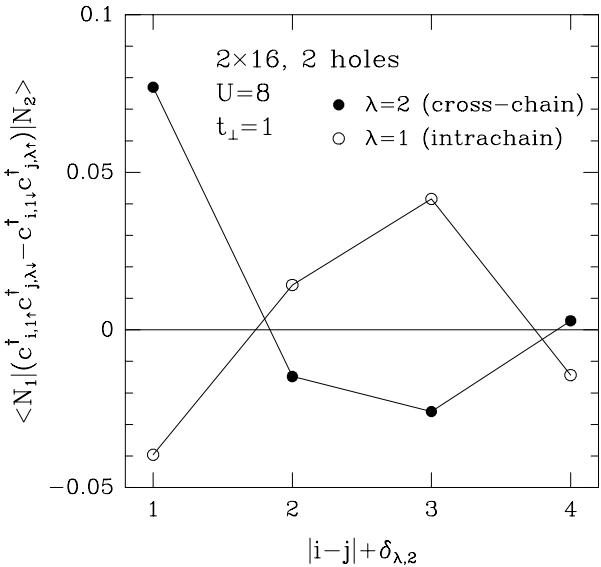, width=8.3cm}
\end{center}
\caption{
The matrix element 
$\langle N_2 |
(c^\dagger_{{\bf r},\uparrow} c^\dagger_{{\bf r'},\downarrow} - 
c^\dagger_{{\bf r},\downarrow} c^\dagger_{{\bf r'},\uparrow})
| N_1 \rangle$
of the pair creation operator between a state 
$|N_1\rangle$ with two holes and $|N_2\rangle$ with four holes on a 
$2\times 16$ 
lattice with $U=8$ and $t_\perp=1.0$.
The distance $\ell$ is the rung separation $|i-j|$ for intrachain part
and $|i-j| +1$ for the cross--chain part.
}
\label{figpairwf}
\end{figure}

In order to understand further the nature of the interaction between holes in
the antiferromagnetic spin liquid, we have examined the pair wave
function of two holes directly.
We do this by calculating the matrix element of the pair creation operator, 
\begin{equation}
\langle N_2 | \Delta^\dagger_{\bf rr'} | N_1 \rangle= 
\langle N_2 |
(c^\dagger_{{\bf r},\uparrow} c^\dagger_{{\bf r'},\downarrow} - 
c^\dagger_{{\bf r},\downarrow} c^\dagger_{{\bf r'},\uparrow})
| N_1 \rangle
\end{equation}
between a state $|N_1\rangle$ with $N$ spin up and $N$ spin down
electrons, and a state $|N_2\rangle$ with $N-1$ spin up and $N-1$ spin
down electrons.
Here the index ${\bf r} = (i,\lambda)$ denotes the position.
The pair wave function can then be extracted by examining the
dependence of this matrix element on $({\bf r}-{\bf r'})$.
The result for the matrix element between the state with four holes
($|N_1\rangle$) and the state with two holes ($|N_2\rangle$)
at $t_\perp=1$ ($N=14$ on a $2\times 16$ lattice) 
is shown in Fig.\ \ref{figpairwf}.
The distance along the ordinate is the rung separation plus the
cross--chain distance (0 or 1), so that a value of 1 corresponds to
nearest--neighbor points either along or between the chains, and
subsequent points are further along the chains.
We have also averaged the matrix element for a number of site pairs with
the same $({\bf r}-{\bf r'})$ in order to reduce the finite--size effects
due to the open boundaries.

\vspace*{-0.8cm}

As can be seen in the plot, the largest amplitude for the pair
matrix element is for a nearest--neighbor site across a rung, and the
second largest for a nearest--neighbor site on the same chain.
The two amplitudes have opposite signs, indicating a d--wave--like
symmetry.
The d--wave--like structure of the pair wave function is maintained
as the separation is increased along the chain.
In addition, the amplitude is strongly suppressed when the two
particles are created on sites
on the same sublattices, as can be seen by the smaller amplitude at
every other point.
This structure can be understood in a strong--coupling picture by
considering a dimer RVB state\cite{rvbprl},
which gives a good qualitative 
description of the two--chain Heisenberg system.
Here, we have added two holes to a half--filled system that is close
to the Heisenberg limit.
When two holes are added on the same sublattice within the dimer RVB
state, the RVB state is frustrated in that the number of possible
valence bond configurations is greatly reduced, leading to a higher
energy.
Therefore, the amplitude of the state in which both holes are on the
same sublattice is suppressed in
order to minimize the probability of the system being in such a
configuration.

We have also examined the pair wave function by calculating similar
matrix elements for $t_\perp = 2.0$ and $t_\perp=0.3$ in order to
examine the large $t_\perp$ Luttinger liquid (C1S1) phase and the
small $t_\perp$ phase. 
We find that the pair wave function does not maintain the coherent
structure shown in Fig.\ \ref{figpairwf} for separations larger than the
nearest--neighbor separations, indicating that this structure is only
present in the C1S0 part of the phase diagram.

\begin{figure}
\begin{center}
\epsfig{file=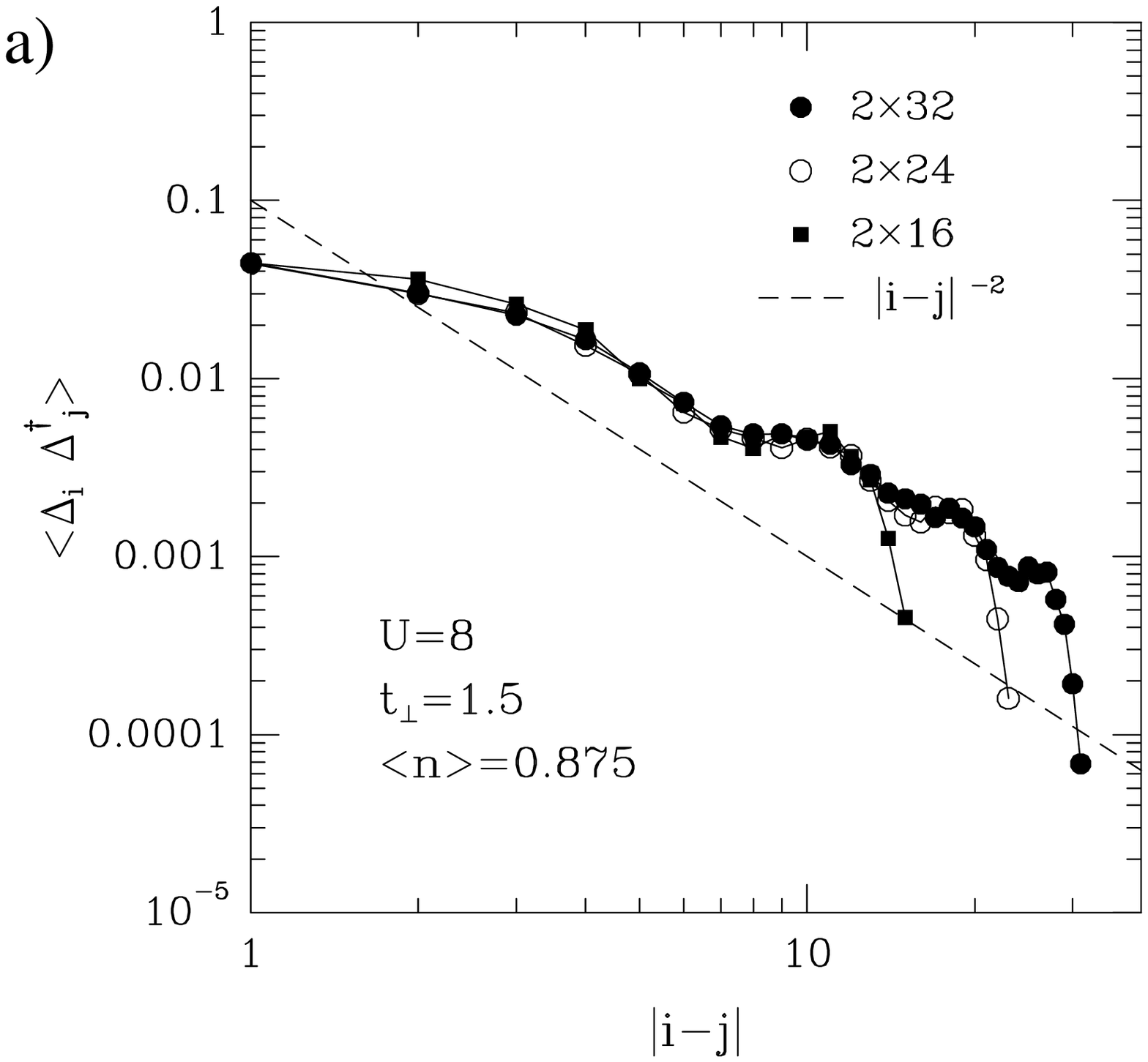, width=8.3cm}
\epsfig{file=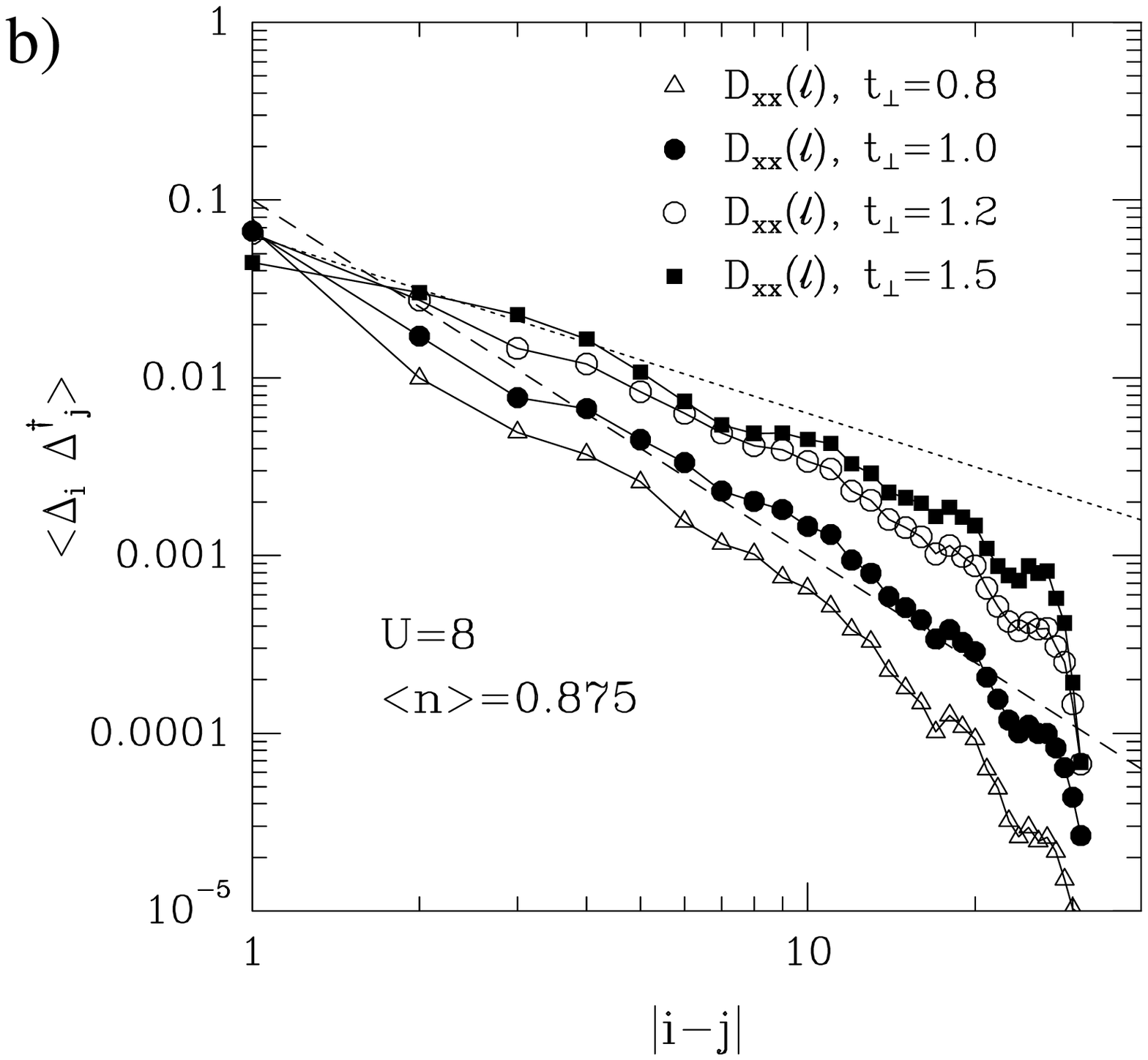, width=8.3cm}
\end{center}
\caption{
The pair--pair correlation function 
$\langle \Delta(i) \Delta^\dagger(j) \rangle$
plotted on a log--log scale for
for $U=8$ and $\langle n \rangle = 0.875$.
In (a) the $t_\perp=1.5$ susceptibility is plotted for different
system sizes, and in (b) the susceptibility is plotted for various
$t_\perp$ on a $2\times 32$ lattice.
The dashed line gives the $|i-j|^{-2}$ decay of the $U=0$ pair
susceptibility, and the short dashed line in (b) shows a decay of
$|i-j|^{-1}$.
}
\label{figpaircorr}
\end{figure}

We have now ascertained that there is a tendency towards pairing within
the doped, spin gap phase (C1S0), and have examined the shape of the
pair wave function.
However, in order to determine the nature of the ground state,
particularly with a view towards possible ordering via
three--dimensional crossover in the real materials, we must examine
the long--distance behavior of the correlation functions.
We choose a pairing order parameter that creates a pair locally in
real space, but has a large overlap with the actual pair wave function:
\begin{equation}
\Delta^\dagger(i) = 
c^\dagger_{i,1,\uparrow} c^\dagger_{i,2\downarrow} - 
c^\dagger_{i,1,\downarrow} c^\dagger_{i,2\uparrow}.
\end{equation}
This operator creates a pair in a spin singlet on two neighboring
sites on a rung (labeled $i$).
We then calculate the corresponding pair correlation function,
$\langle \Delta(i) \Delta^\dagger(j) \rangle$,
as a function of distance between the rungs, $|i-j|$.
This pair correlation is shown, plotted on a log--log scale for
various system sizes and $t_\perp$ values within the spin--gapped
phase at $\langle n \rangle = 0.875$ in Fig.\ \ref{figpaircorr}.

In order to exhibit the finite--size effects, we show
the pair correlation function on $2\times 8$, $2\times 16$, and
$2\times 32$ lattices in Fig.\ \ref{figpaircorr}(a).
Recall that the local density, as seen in Fig.\ \ref{figholeden},
can have substantial dependence on lattice position.
Since the correlation functions on the finite lattice with open
boundaries are not translationally invariant, we must choose
one or more $i$, $j$ pairs in order to best approximate infinite
system behavior.
Originally, we calculated the correlation functions by choosing one
$i$, $j$ pair for each $|i-j|$ so that the pair was as symmetrical
about the center of the lattice as possible.
For odd separations, $i$ and $j$ can be chosen symmetrically about
the lattice center (for a lattice of even length), whereas for even
separations, they cannot be.
Such an odd--even effect produced, for example, a spurious peak at 
${\bf q} = (\pi,\pi)$ in the Fourier transform of the spin--spin
correlation function for the doped system (shown in Fig.\ \ref{figSq}).
In general, correlation functions in real space calculated in this way
show spurious oscillations which can be seen by comparing the correlation
functions at a given distance on different lattice sizes.
In order to try to remove these effects, we average correlation
functions over a number of $i$, $j$ pairs for each $|i-j|$.
We have found it best to take enough pairs so that one averages over
the wavelength of the oscillations in the local quantities.
Here we average over six pairs, starting with the symmetrically
placed pair and then proceeding down the lattice.
When $|i-j|$ gets near the lattice size, the number of possible
$i$, $j$ pairs is smaller due to the proximity to the boundaries.
As seen in Fig.\ \ref{figpaircorr}(a), the pair correlation function
treated in this way shows only small finite--size effects until the
last few points before the boundary are reached.
There are oscillations remaining in the correlation function, but
these are not a finite--size effect and are present in the
thermodynamic limit.
The unphysical peaks at ${\bf q}= (\pi,\pi)$ for the
doped cases in Fig.\ \ref{figSq} disappeared after this averaging
procedure was applied to $S(i,j,\lambda)$.

\vspace*{-0.8cm}

In Fig.\ \ref{figpaircorr}(b), we plot the pair correlation function 
$\langle \Delta(i) \Delta^\dagger(j) \rangle$ on a log--log
scale as a function of 
$|i-j|$ on a $2\times 32$ lattice for a variety of $t_\perp$ values 
within the C1S0 phase at $\langle n \rangle = 0.875$.
Also shown on the plot are lines representing the power law decays
$|i-j|^{-1}$ and $|i-j|^{-2}$.
As can be seen, the pair correlation functions are approximately
straight lines on the log--log scale, implying a power--law decay of
the correlations.
If this power law has the form $|i-j|^{-\nu}$,
$\nu$ ranges from 
slightly greater than $2$ for $t_\perp=0.8$ to approximately $1$ for
$t_\perp=1.5$.
In the noninteracting, $U=0$, system, the pairing correlation function
decays as $|i-j|^{-2}$, so the pairing is only enhanced over that in the
noninteracting system for $t_\perp > 1.0$.
The strength of the pairing correlations is correlated with the size
of the spin gap, as shown in Fig.\ \ref{figgapinf}.
The average strength of the pair correlation function, calculated by
averaging $\langle \Delta(i) \Delta^\dagger(j) \rangle$ between $|i-j|=8$ and 
$|i-j|=12$ lattice spacings, is shown as a dashed line.
In the isotropic case, the pairing exponent
$\nu \approx 2$.
Since this phase is predicted to be a C1S0 phase by the BFRG, 
we would expect $\nu = 1/K_\rho$ within a Luther--Emery picture.
Recall that the weakest criterium for dominant pair correlations within
a Luther--Emery picture is that $K_\rho > 1$, i.e. that $\nu < 1$.
Therefore, the pairing correlations would not be dominant for the
isotropic system within a Luther--Emery picture, and a direct
application of this model to the La$_{1-x}$Sr$_x$CuO$_{2.5}$ materials
would not predict a superconducting ground state.

\vspace*{-0.8cm}
\begin{figure}
\begin{center}
\epsfig{file=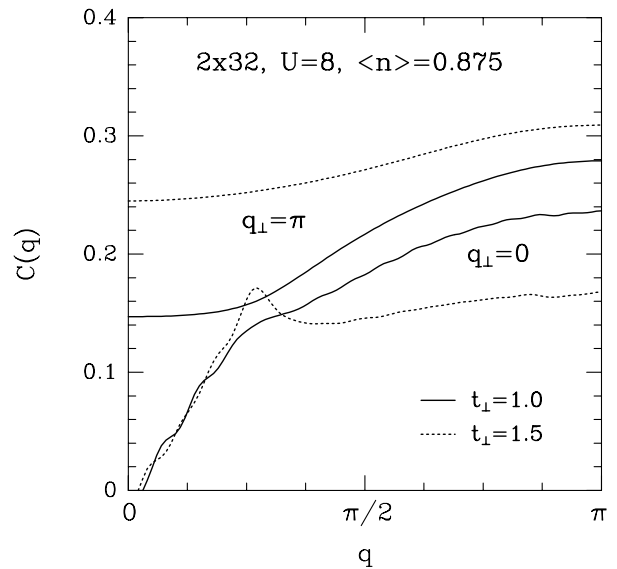, width=8.3cm}
\end{center}
\caption{
The continuous Fourier transform of the density--density correlation
function $C({\bf q})$ on a $2 \times 32$ lattice
with $U=8$ and $\langle n \rangle = 0.875$ for $t_\perp=1.0$ and
$t_\perp=1.5$.
The upper curves are the $q_\perp=\pi$ branches and the lower curves
the $q_\perp=0$ branches.
}
\label{figcdwq}
\end{figure}

We now search for a CDW correlation function 
with a reciprocal behavior to the pairing
correlations, as in Eq.\ \ref{eqle}.
In Fig.\ \ref{figcdwq}, we plot the continuous Fourier transform of
the CDW correlation function
\begin{equation}
C(i,j,\lambda) = 
\langle n_{i,\lambda} n_{j,1} \rangle 
- \langle n_{i,\lambda}\rangle \langle n_{i,1}\rangle
\label{eqdenden}
\end{equation}
at $\langle n \rangle =0.875$, for $t_\perp=1.0$ and $t_\perp=1.5$,
within the C1S0 spin--liquid phase.
The real--space correlation function is averaged and
fourier--transformed in the same way as the spin--spin correlation
function of Eq.\ \ref{eqSij}.
As can be seen, there is no distinct peak at the ``$2 k_F$'' wave vector
${\bf q} = (q^*,\pi)$, where 
$q^* = k_f^b + k_F^a = \pi \langle n \rangle$.
However, there is a peak at ${\bf q} = (\pi/4,0)$ which grows and
sharpens with increasing $t_\perp$.
This wavevector is the correct value for the ``$4k_F$'' CDW
correlations, for which $|2(k_F^b + k_F^a) ~ \text{mod} ~ 2\pi| = \pi/4$.
Within a bosonization treatment\cite{nagaosa,balents}, the 
``$4 k_F$'' peak should not appear in the density--density
correlation function of Eq.\ \ref{eqdenden}, but only in a
density--density correlation function composed of four fermion
densities.
We believe that the fact that it appears in $C({\bf q})$ is due to the
band curvature, which is not included in the bosonization picture, and
effectively includes higher order density--density correlations
in $C({\bf q})$.

\begin{figure}
\begin{center}
\epsfig{file=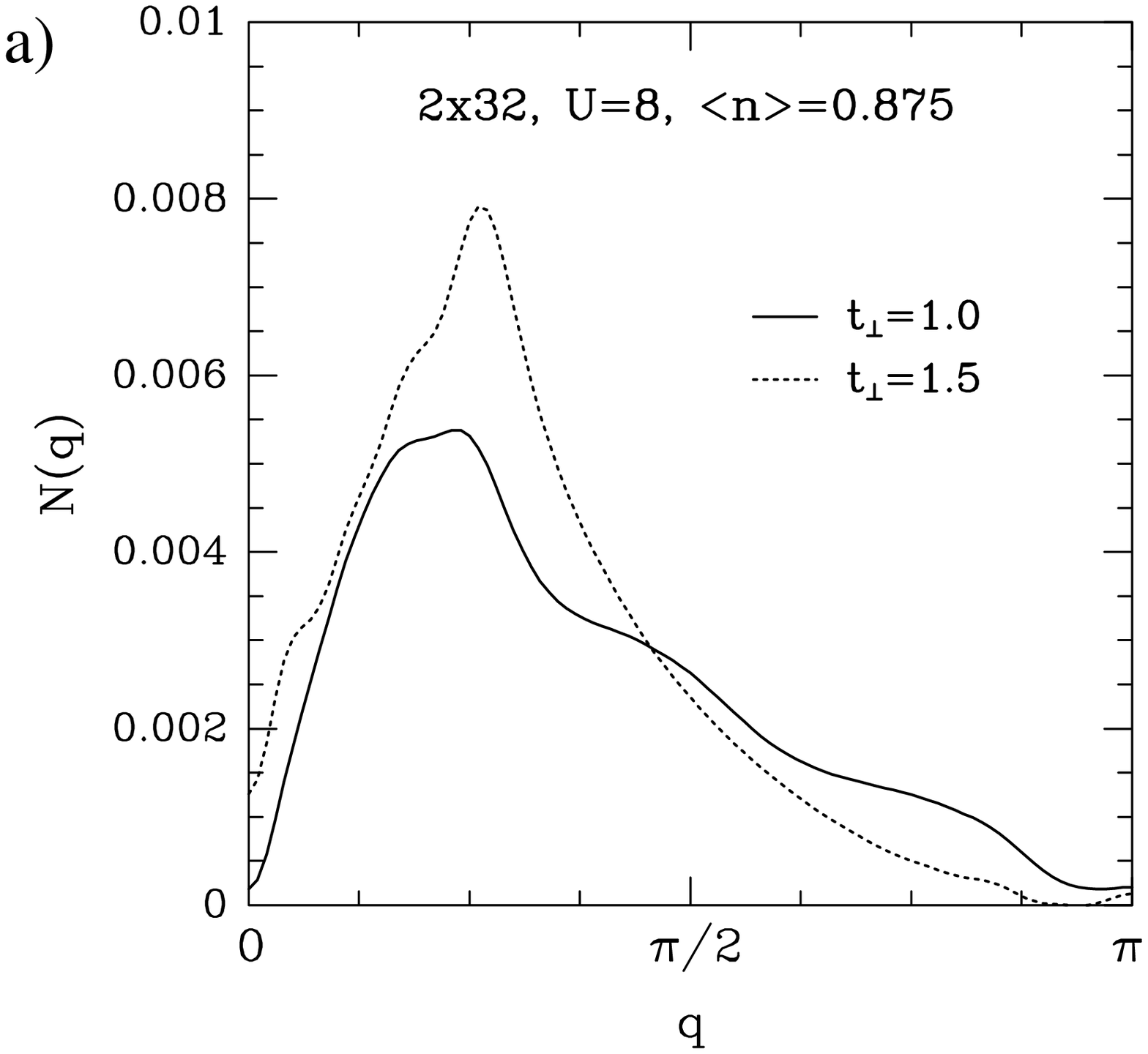, width=8.3cm}
\epsfig{file=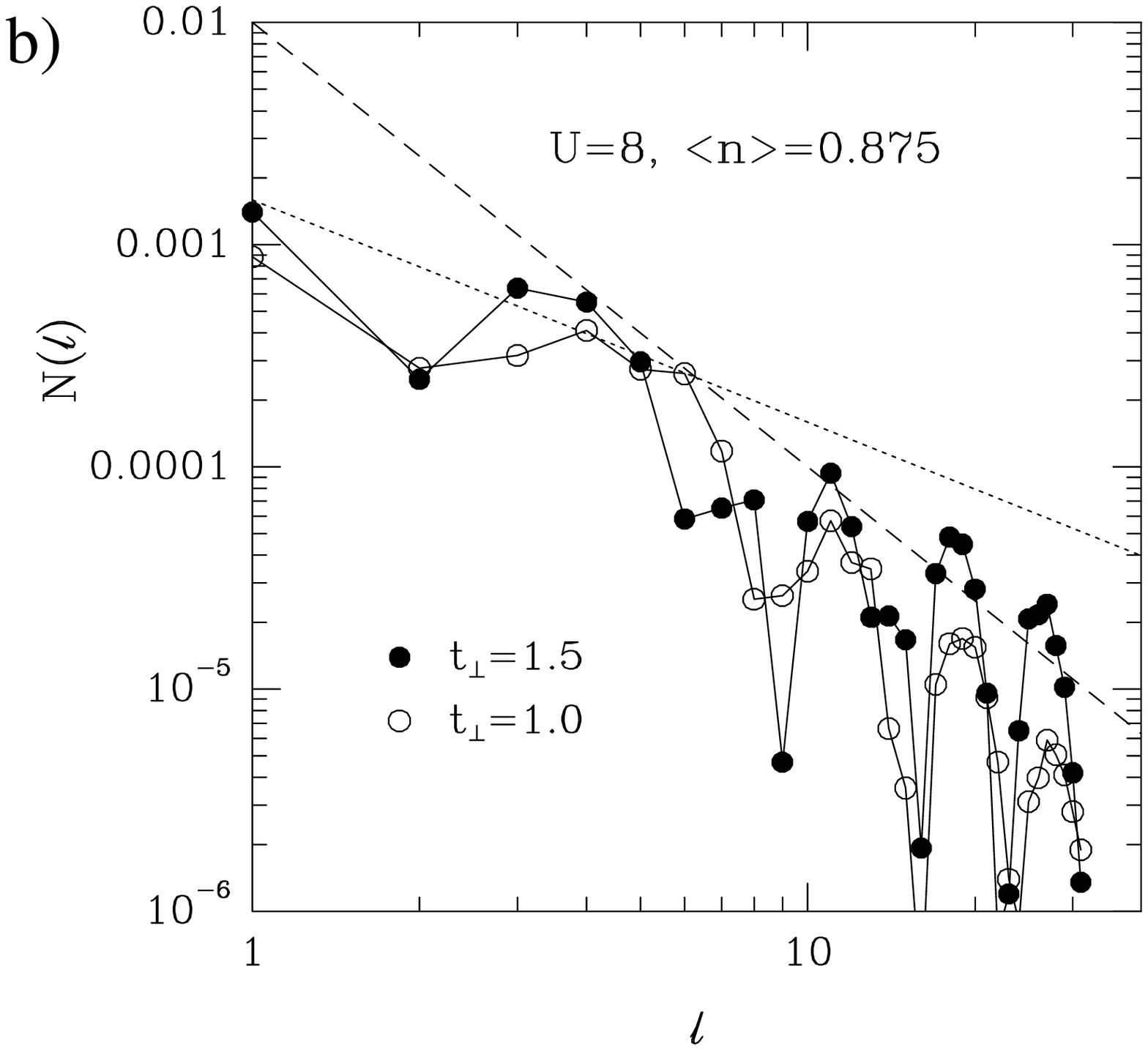, width=8.3cm}
\end{center}
\caption{
The $4k_f$ density--density correlation function $N(i,j)$, as defined
in Eq.\ (13), (a) fourier--transformed and (b) in real space on a 
$2 \times 32$ lattice for $U=8$, $\langle n \rangle =0.875$, and
$t_\perp=1.0$, 1.5.
}
\label{fignag}
\end{figure}

We have isolated the ``$4k_F$'' portion of the density--density
correlation function by calculating the correlation function of a
simple effective rung density
\begin{equation}
n^{\rm eff}_i = n_{i,1\uparrow} n_{i,2\downarrow}.
\end{equation}
Since $n^{\rm eff}_i$ is composed of four fermion operators, rather
than two as in the ordinary density--density correlation function, in
order to form the correlation function, 
$\langle n^{\rm eff}_i n^{\rm eff}_j \rangle$
one must subtract off fourteen different
disconnected pieces, not only
$\langle n^{\rm eff}_i \rangle \langle n^{\rm eff}_j \rangle$, as in
Eq.\ \ref{eqdenden}.
We will not write all the disconnected terms explicitly here.
However, the subtraction can be carried out to obtain the real space
correlation 
function $N(i,j)$ which can be averaged over a number of $i$, $j$
pairs and fourier--transformed with respect to the rung index, as was
done for the spin--spin and density--density correlation functions
above, to obtain $N(q)$.
The fourier--transformed rung--density correlation function $N(q)$ is
plotted for $t_\perp=1.0$ and 
$t_\perp=1.5$ in Fig.\ \ref{fignag}(a).
As can be seen, the ``$4 k_f$'' peak at $q=\pi/4$ is now well--defined,
and the broad background seen in $C({\bf q})$ is no longer present.
As in Fig.\ \ref{figcdwq}, the peak at $q=\pi/4$ grows in size and sharpness
as $t_\perp$ is increased from 1.0 to 1.5.
In Fig.\ \ref{fignag}(b) we exhibit the averaged real--space
correlation function $N(\ell)$, where $\ell=|i-j|$, on a log--log scale
in order to estimate the exponent of the decay.
The envelope of the correlations is consistent with a straight line on
the log--log scale and therefore an algebraic decay, although the 
``$4 k_F$'' structure makes it difficult to directly determine how the
envelope decays.
If we assume a form $|i-j|^{-\gamma}$ for the decay of the envelope, 
one can estimate $\gamma$ to be slightly greater than $2$ for
$t_\perp=1.0$, and slightly less than $2$ for $t_\perp=1.5$.

Therefore, the behavior is opposite to what one would expect from the
Luther--Emery model: both the pairing exponent $\nu$ and the 
``$4 k_F$'' CDW exponent $\gamma$ decrease as 
$t_\perp$ increases within the doped spin--liquid phase, and clearly
do not obey the relationship $\nu \gamma = 1$.
It is not clear why the exponents do not behave as expected.
One possibility is that the identification of this phase with the
Luther--Emery--like phase found in the bosonization calculations is
incorrect.
Another possibility is that there is another CDW--like correlation
function that behaves in the expected way.
A fact that could also be important is that the CDW should be quite
slowly decaying.
For example, in the isotropic case, $\nu \approx 2$
so that $\gamma$  should be $\approx 1/2$.
Perhaps such long--range CDW correlations effectively put the system
into an ordered CDW state on a finite lattice with
open boundaries. 
In this case, the ``$4k_F$'' CDW correlation
function, which measures the decay of the fluctuations, might not
behave in the expected way.

\vspace*{-0.8cm}

\section{\label{CON} CONCLUSION}

We have explored the ground state phase diagram of the two--leg
Hubbard ladder as a function of the band filling $\langle n \rangle$ 
and the interchain hopping matrix element $t_\perp$, concentrating on
the case of intermediate to strong on--site interaction, primarily
$U=8$.
For the half--filled system, we find a spin--gapped insulating phase
for all $t_\perp$ and $U$ accessible to our numerical density--matrix
renormalization group calculations.
For weak $U$, there is a relatively sharp transition from a
spin--liquid insulator at small $t_\perp$ to a band insulator for
$t_\perp > 2.0$.
As $U$ is increased, this transition becomes a gradual crossover more
consistent with a Heisenberg--model picture.
However, at large enough $t_\perp$, we would expect a return to a band
insulator for all finite $U$.
This behavior is consistent with recent weak--coupling renormalization
group and bosonization calculations \cite{balents}, as well as,
in the large $U$ limit, strong--coupling Heisenberg model
calculations\cite{gopalan,barnes}.

When the half--filled insulator is doped with holes, we find a number
of distinct phases.
The major feature dominating the phase diagram is associated with the
weak--coupling one--band to two--band transition.
For $t_\perp$ larger than the $U=0$ band transition value, 
$t_{\perp c}$, our numerical results show behavior consist with that
of a Luttinger liquid, whose band filling corresponds to the filling
of the $U=0$ bonding band.
The development of a charge gap, expected in the half--filled
Luttinger liquid, clearly occurs at quarter filling in the two--leg
ladder for $t_\perp > t_{\perp c}$.
For $t_\perp < t_{\perp c}$, there is a region near $t_{\perp c}$ in
which the spin gap still vanishes.
This is consistent with a weak--coupling renormalization group (RG)
calculation of Balents and Fisher\cite{balents}, which predicts phases
with gapless spin 
excitations where the bands overlap, but the Fermi velocities are
very different.
In addition, for light doping and small $t_\perp$, we find the spin
gap vanishes at a finite $t_\perp$ consistent with a half--filled
bonding band, a point at which the Balents--Fisher calculation predicts a
phase with a vanishing spin gap associated with umklapp processes in
the half--filled bonding band.
For the remainder of the $t_\perp < t_{\perp c}$ region, and
especially in the light to moderately doped isotropic case 
there is a spin--gapped phase
with gapless charge excitations in which there are algebraically
decaying $d$--wave--like superconducting correlations, and ``$4k_F$''
CDW correlations.
However, the pairing correlations do not seem to be dominant in the
isotropic system, and exponents of the CDW and pairing
correlations do not seem to obey the inverse relationship predicted by a
bosonization picture\cite{nagaosa,balents}.

\section*{ACKNOWLEDGMENTS}

S.R.W and R.M.N. acknowledge support from the Office of Naval Research under
grant No. N00014-91-J-1143.
S.R.W. and D.J.S. acknowledge support
from the NSF under Grants No.\ DMR-9509945 and No.\ DMR 92-2507,
respectively.
The calculations were performed on a Cray YMP and a Cray C90 at the
San Diego Supercomputer Center.


\end{document}